# IN CLASSICAL MECHANICS OBJECTIVITY LOST WHEN RIEMANN-LIOUVILLE OR CAPUTO FRACTIONAL ORDER DERIVATIVE ARE USED

Agneta M. Balint[1] and Stefan Balint[2]

[1]Department of Physics, West University of Timisoara, Bulv. V. Parvan 4, 300223 Timisoara, Romania

[2]Department of Computer Science, West University of Timisoara, Bulv. V. Parvan 4, 300223 Timisoara, Romania.

Corresponding author: agneta.balint@e-uvt.ro; stefan.balint@e-uvt.ro

**Abstract.** In section.1 the objectivity in science is presented shortly. In section.2 some details concerning the objectivity in the case of the mechanical movement description of a material particle are given. In section.3 details concerning the objectivity of the description in the Newtonian continuum mechanics are given. The Riemann-Liouville and the Caputo fractional order derivatives are presented shortly in section 4. In section 5 the lost of the objectivity of the mechanical movement descriptions, presented in sections 2,3, due to the use of the Riemann-Liouville or the Caputo fractional order derivatives is presented. This is followed, in section 6, by the presentation of some recent papers which propose the use of these fractional order derivatives, instead of the integer order derivatives, in the description of some physical phenomena. It is underlined that the objectivity in these last papers is ignored. At the end of the present paper instead of conclusion we formulate the following question: if a mathematical description of a real phenomenon is not objective then what is the interpretation of the reported results and how this results have to be used?

## 1. Objectivity in science

The concept of objectivity in science means that qualitative and quantitative descriptions of phenomena remain unchanged when the phenomena are observed by different observers; that is, it is possible to reconcile observations of the process into a single coherent description of it. This type of requirement concerning the mechanical events descriptions is originated in the following statements: Galileo Galilee (1564-1642) said" The mechanical event is independent from the observer. For frames moving uniformly with respect to each other, both states are mechanically equivalent."; Isaac Newton (1643-1727) said " The mechanical event is independent from the observer. This holds also for accelerated systems if the frames of references are fixed with respect to absolute space (with respect to the fixed stars)"; Albert Einstein (1879-1955) said" The mechanical event is independent from the observer. There is no special reference point. The same holds for accelerated systems (general relativity).Even more, if the theory is subjected to relativity, it should be generally covariant under all transformations, not just rigid body motions." [1].

## 2. Objectivity of the description of the movement of a material particle in Newtonian mechanics.

In the following, different details of the objectivity is illustrated in the case of the description of the movement of a material particle [2].

*2.1. Objectivity of the description of the position of a material particle.*

In classical mechanics one observer $O$ represents a material particle by a point $P$ called material point in the three-dimensional affine Euclidian space $E_3$. For describe the movement of the material point observer $O$ choose a fixed orthogonal reference frame $R_O = (O; \vec{e}_1, \vec{e}_2, \vec{e}_3)$ in the space $E_3$, a moment of time $M_O$ for fixing the origin of the time measurement and a unit [second] for the time measuring. For the observer $O$, a moment of time $M$ which is earlier than $M_O$ is represented by a negative real number $t_M < 0$, a moment of time $M$ which is later than $M_O$ is represented by a positive real number $t_M > 0$ and the moment of time $M_O$ is represented by the real number $t_{M_O} = 0$. At an arbitrary moment of time $M$, represented by $t_M$, the observer considers the coordinate $(X_1(t_M), X_2(t_M), X_3(t_M))$ of the material point at the moment of time $t_M$ with respect to the reference frame $R_O$, and describes the position of the material point with the set of the real functions $X_1(t_M), X_2(t_M), X_3(t_M)$.

A second observer $O^*$ uses a similar procedure and describes the position of the same material point with the set of the real functions $X^*_1(t_M), X^*_2(t_M), X^*_3(t_M)$, representing the coordinates of the material point with respect to a second fixed orthogonal reference frame $R_{O^*} = (O^*; \vec{e}^*_1, \vec{e}^*_2, \vec{e}^*_3)$. For the observer $O^*$ the origin of the time measurement is $M_{O^*}$, the unit is [second] and a moment of time $M$ which is earlier than $M_{O^*}$ is represented by a negative number $t^*_M < 0$, the moment of time $M_{O^*}$ is represented by the real number $t^*_{M_{O^*}} = 0$ and a moment of time $M$ which is later than $M_{O^*}$ is represented by a positive real number $t^*_M > 0$.

In case of the observer $O$ a moment of time $M$ is described by the real number $t_M$ and in case of the observer $O^*$ by the real number $t^*_M$. For the numbers $t_M$ and $t^*_M$ the following relations hold:

$$t_M = t^*_M + t_{M_{O^*}} \tag{1}$$

$$t^*_M = t_M + t^*_{M_O} \tag{2}$$

In the above relations $t_{M_{O^*}}$ is the real number which represents the moment $M_{O^*}$ in the system of time measuring of the observer $O$ and $t^*_{M_O}$ is the real number which represents the moment $M_O$ in the system of time measuring of the observer $O^*$.

For an arbitrary moment of time $M$, the coordinates $(X_1(t_M), X_2(t_M), X_3(t_M))$ in $R_O$ and $X^*_1(t_M), X^*_2(t_M), X^*_3(t_M)$ in $R_{O^*}$ represent points in the three dimensional affine Euclidian space $E_3$. These points have to coincide with the material point position at the moment of time $M$. Therefore, for the coordinates the following relations hold:

$$X_k(t_M) = X_{kO^*} + a_{1k} X^*_1(t^*_M) + a_{2k} X^*_2(t^*_M) + a_{3k} X^*_3(t^*_M) \quad k = 1,2,3$$
$$t^*_M = t_M + t^*_{M_O} \tag{3}$$

or equivalently

$$X^*_k(t^*_M) = X^*_{kO} + a_{k1} X_1(t_M) + a_{k2} X_2(t_M) + a_{k3} X_3(t_M) \quad k = 1,2,3$$
$$t^*_M = t_M + t^*_{M_O} \tag{4}$$

The significance of the quantities appearing In the above relations are:

$a_{ij} = \langle \vec{e}^*_i, \vec{e}_j \rangle$ = constant = scalar product of the unit vectors $\vec{e}^*_i$ and $\vec{e}_j$ in $E_3$.

$(X_{1O^*}, X_{2O^*}, X_{3O^*})$ are the coordinates of the point $O^*$ in the reference frame $R_O$

$(X^*_{1O}, X^*_{2O}, X^*_{3O})$ are the coordinates of the point $O$ in the reference frame $R_{O^*}$

Relations (3) or (4) reconcile description of the position made by the two observers and make possible the description of the position by the set of functions $X_1(t_M), X_2(t_M), X_3(t_M)$ or by the set of functions $X^*_1(t_M), X^*_2(t_M), X^*_3(t_M)$.

## 2.2. Objectivity of the velocity concept of the material point

In the above type of description of the position, the concept „velocity $\vec{V}_M$ at the moment $M$ of the material particle" is objective. In order to see what this means exactly, remember that in the description $O$ „velocity $\vec{V}_M$ at the moment $M$ of the material particle „ is the vector in $E_3$ obtained translating the vector $\vec{X}'(t_M) = X'_1(t_M)\vec{e}_1 + X'_2(t_M)\vec{e}_2 + X'_3(t_M)\vec{e}_3$ in the point of coordinates $(X_1(t_M), X_2(t_M), X_3(t_M))$. Here $X'_k(t_M)$ is the first order derivative of the function $X_k(t_M)$ at $t_M$ for $k = 1,2,3$. In the description of $O^*$ „velocity $\vec{V}^*_M$ at the moment $M$ of the material particle „ is the vector in $E_3$ obtained translating the vector $\vec{X}^{*'}(t^*_M) = X^{*'}_1(t^*_M)\vec{e}^*_1 + X^{*'}_2(t^*_M)\vec{e}^*_2 + X^{*'}_3(t^*_M)\vec{e}^*_3$ in the point of coordinates $X^*_1(t^*_M), X^*_2(t^*_M), X^*_3(t^*_M)$. Here $X^{*'}_k(t^*_M)$ is the first order derivative of the function $X^*_k(t^*_M)$ at $t^*_M$ for $k = 1,2,3$.

Objectivity of the concept „ velocity at the moment $M$ of the material point means $\vec{V}_M = \vec{V}^*_M$ in $E_3$, i.e. the velocity of the material point is independent on observer. Equality $\vec{V}_M = \vec{V}^*_M$ in $E_3$ can be obtained by differentiating in (3) and obtaining:

$$X'_k(t_M) = a_{1k} X^{*'}_1(t^*_M) + a_{2k} X^{*'}_2(t^*_M) + a_{3k} X^{*'}_3(t^*_M) \quad k = 1,2,3 \tag{5}$$

Equations (5) shows that $\vec{V}_M = \vec{V}^*_M$ in $E_3$.

## 2.3. Objectivity of the acceleration concept of the material point

By using the observer $O$ description the concept „acceleration $\vec{A}_M$ at the moment of time $M$ of the material point" is the vector in $E_3$ obtained translating the vector $\vec{X}''(t_M) = X''_1(t_M)\vec{e}_1 + X''_2(t_M)\vec{e}_2 + X''_3(t_M)\vec{e}_3$ in the point of coordinates $(X_1(t_M), X_2(t_M), X_3(t_M))$. Here $X''_k(t_M)$ is the second order derivative of the function $X_k(t_M)$ at $t_M$ for $k = 1,2,3$.

Using observer $O^*$ description the concept „acceleration $\vec{A}^*_M$ at the moment of time $M$ of the material point" is defined similarly.

Objectivity of the concept „acceleration at the moment $M$ of the material point" means $\vec{A}_M = \vec{A}^*_M$ in $E_3$, i.e. the concept is independent on observer. Equality $\vec{A}_M = \vec{A}^*_M$ in $E_3$ can be obtained by differentiating twice in (3) and obtaining:

$$X''_k(t_M) = a_{1k} X^{*''}_1(t^*_M) + a_{2k} X^{*''}_2(t^*_M) + a_{3k} X^{*''}_3(t^*_M) \qquad k=1,2,3 \qquad (6)$$

Equations (6) show that $\vec{A}_M = \vec{A}^*_M$ in $E_3$.

### 2.4. Objectivity of the equation describing the dynamics of the material point in Newtonian mechanics.

The second law of Newton „the rate of change of momentum is proportional to the impressed force, and takes place in the direction of the straight line in which the force acts" [2] in terms of the description of observer $O$, leads to the conclusion that the functions $X_1(t_M), X_2(t_M), X_3(t_M)$, describing the motion, satisfy the following system of differential equations:

$$mX''_k(t_M) = F_{kO}(t_M, X_1(t_M), X_2(t_M), X_3(t_M), X'_1(t_M), X'_2(t_M), X'_3(t_M)) \quad k=1,2,3 \qquad (7)$$

Here: $m$ represents the material point mass, $X_k(t_M)$, $k=1,2,3$ represent the coordinates of the position of the material point in $R_O$, $X'_k(t_M)$, $k=1,2,3$ represent the velocity components of the material point in $R_O$, $X''_k(t_M)$, $k=1,2,3$ represent the acceleration components of the material point and $F_{kO}$, $k=1,2,3$ represents the components (in the reference frame $R_O$) of the force field $\vec{F}$ acting on the material point.

In terms of the description of observer $O^*$, the same law of Newton, leads to the conclusion that the functions $X^*_1(t^*_M), X^*_2(t^*_M), X^*_3(t^*_M)$ describing the motion under the action of the same force field $\vec{F}$, in the reference frame $R_{O^*}$, satisfy the following system of differential equations:

$$mX^{*''}_k(t^*_M) =$$
$$F^*_{kO^*}(t^*_M, X^*_1(t^*_M), X^*_2(t^*_M), X^*_3(t^*_M), X^{*'}_1(t^*_M), X^{*'}_2(t^*_M), X^{*'}_3(t^*_M)) \qquad (8)$$
$$k=1,2,3$$

Note that in (8) $F^*_{kO^*}$ represents the components of the same force field $\vec{F}$ with respect to $R_{O^*}$ and satisfies:

$$F^*_{kO^*} = a_{k1} F_{1O} + a_{k2} F_{2O} + a_{k3} F_{3O} \qquad k=1,2,3 \qquad (9)$$

The systems of differential equations are different, but their solution describes the movement of the material point under the action of the same force field. This can be proven showing that if $X_1(t_M), X_2(t_M), X_3(t_M)$ is a solution of (7), then the functions $X^*_1(t^*_M), X^*_2(t^*_M), X^*_3(t^*_M)$, given by (4), is a solution of (8) and vice versa.

In other words, the dynamics of the material point can be described by the system (7) or by the system (8). This means that the system of differential equations (7) and (8) is independent on the observer. Each of them can be considered the system of differential equations which describes the dynamics of the material point.

### 3. Objectivity in Newtonian continuum mechanics.

In the following, different details of the objectivity is illustrated in the case of the description used in Newtonian continuum mechanics [3].

*3.1 Objectivity of the description of the movement of a continuum material body.*

In the classical theory of elasticity [3] an observer represent each particle of a material body $B$ by a point in $E_3$ called material point. More precisely, at a moment of time $M$ a particle $P$ of the material body $B$ is represented by a material point $P_M$ in $E_3$. So, at the moment of time $M$ the material body is represented by a subset $S_M$ of material points of $E_3$. Observer describes the movement of the material body describing the movement in $E_3$ of each material point of $S_M$.

This description is made by observer $O$ with functions of the form:

$$Y_k = Y_k(t_M, X_1, X_2, X_3) \qquad k = 1,2,3 \text{ and } (X_1, X_2, X_3) \in S_{M_O} \tag{10}$$

where: $(X_1, X_2, X_3)$ are the coordinates with respect to $R_O$ of an arbitrary material point $P_{M_O}$ from $S_{M_O}$ at the moment $M_O$ (i.e. $t_M = t_{M_O} = 0$), and $(Y_1(t_M, X_1, X_2, X_3), Y_2(t_M, X_1, X_2, X_3), Y_3(t_M, X_1, X_2, X_3))$ are the coordinates with respect to $R_O$ of the same material point $P_{M_O}$ at the moment of time $M$ (i.e. $t_M > 0$). An obvious property of the functions appearing in the description (10) is that they satisfy equalities $Y_k(0, X_1, X_2, X_3) = X_k$ for $k = 1,2,3$ and any $(X_1, X_2, X_3) \in S_{M_O}$. In a description of the form (10) is assumed that for any fixed $t_M$ the function $Y_{t_M}(X)$ defined as

$$Y_{t_M}(X_1, X_2, X_3) = (Y_1(t_M, X_1, X_2, X_3), Y_2(t_M, X_1, X_2, X_3), Y_3(t_M, X_1, X_2, X_3)) \tag{11}$$

is bijective from $S_{M_O}$ to $S_M$ as well that the functions $Y_k(t_M, X_1, X_2, X_3)$ are continuously differentiable for $k = 1,2,3$ and the Jacobi matrix of the function $Y_{t_M}(X)$ is nonsingular. Moreover, it is assumed that the inverse function $Y_{t_M}^{-1}$ of $Y_{t_M}$ defined by

$$Y_{t_M}^{-1}(Y_1(t_M, X_1, X_2, X_3), Y_2(t_M, X_1, X_2, X_3), Y_3(t_M, X_1, X_2, X_3)) = (X_1, X_2, X_3) \tag{12}$$

has the same properties.

Observer $O^*$ describes the same movement of the material body with formulas

$$Y^*_k = Y^*_k(t_M^*, X^*_1, X^*_2, X^*_3) \quad \text{for } k = 1,2,3 \text{ and } (X^*_1, X^*_2, X^*_3) \in S_{M_{O^*}} \tag{13}$$

where: $(X^*_1, X^*_2, X^*_3)$ are the coordinates with respect to $R_{O^*}$ of an arbitrary material point $P_{M_{O^*}}$ from $S_{M_{O^*}}$ at $M_{O^*}$ (i.e. $t_{M_{O^*}}^* = 0$) and
$(Y^*_1(t^*, X^*_1, X^*_2, X^*_3), Y^*_2(t^*, X^*_1, X^*_2, X^*_3), Y^*_3(t^*, X^*_1, X^*_2, X^*_3))$ are the coordinates with respect to $R_{O^*}$ of the same material point $P_{M_{O^*}}$ at the moment of time $M$ (i.e. $t^* = t^*_M > 0$).

An obvious property of the functions appearing in the description (13) is that they satisfy $Y^*_k(0, X^*_1, X^*_2, X^*_3) = X^*_k$ for $k = 1,2,3$ and any $(X^*_1, X^*_2, X^*_3) \in S_{M_{O^*}}$. In the description (13) is assumed that for a fixed $t_M^*$ the function $Y_{t_M^*}(X^*)$ defined as

$$Y_{t_M^*}(X^*_1, X^*_2, X^*_3) = (Y^*_1(t_M^*, X^*_1, X^*_2, X^*_3), Y^*_2(t_M^*, X^*_1, X^*_2, X^*_3), Y^*_3(t_M^*, X^*_1, X^*_2, X^*_3)) \quad (14)$$

is bijective from $S_{M_{O^*}}$ to $S_M$ as well that the functions $Y^*_k(t_M^*, X^*_1, X^*_2, X^*_3)$ are continuously differentiable for $k = 1,2,3$ and the Jacobi matrix of the function $Y^*_{t_M^*}(X^*)$ is nonsingular. Moreover, it is assumed that the inverse function $Y^{*-1}_{t_M^*}$ of $Y^*_{t_M^*}$ defined by

$$Y^{*-1}_{t_M^*}(Y^*_1(t_M^*, X^*_1, X^*_2, X^*_3), Y^*_2(t_M^*, X^*_1, X^*_2, X^*_3), Y^*_3(t_M^*, X^*_1, X^*_2, X^*_3)) = (X^*_1, X^*_2, X^*_3) \quad (15)$$

has the same properties.
Relations which reconcile description (10) and (13), made by the two observers, and make possible the description of the movement by one of them, are the followings:

$$X^*_k = X^*_{k0} + \sum_{i=1}^{i=3} a_{ki} \cdot Y_i(t_{M_{O^*}}, X_1, X_2, X_3) \quad \text{for} : k = 1,2,3 \text{ and } (X_1, X_2, X_3) \in S_{M_O} \quad (16)$$

$$Y^*_k(t_M^*, X^*_1, X^*_2, X^*_3) = X^*_{k0} + \sum_{i=1}^{i=3} a_{ki} \cdot Y_i(t_M, X_1, X_2, X_3) \quad \text{for} \quad k = 1,2,3, \quad (17)$$

$$(X_1, X_2, X_3) \in S_{M_O} \text{ and } (X^*_1, X^*_2, X^*_3) \in S_{M_{O^*}}$$

$$t^* = t + t^*_{M_O} \quad (18)$$

*3.2. Objectivity of the concept of displacement vector and that of the Cauchy deformation tensor*

In the theory of deformation by using the observer $O$ description the vector

$$\vec{U} = \sum_{j=1}^{3}(Y_j(t_M, X_1, X_2, X_3) - X_j) \cdot \vec{e}_j \quad (19)$$

translated to the material point $P_{M_O}$ is called the displacement vector of the material point $P_{M_O}$ at the moment of time $M$ (i.e. $t_M = t$) and is denoted by $\vec{D}_M$.

The displacement vector $\vec{D}*_M$ constructed by using observer $O*$ description is obtained translating the vector

$$\vec{U}* = \sum_{j=1}^{3} (Y*_j (t_M*, X*_1, X*_2, X*_3) - X*_j ) \cdot \vec{e}*_j \qquad (20)$$

to the material point $P_{M_O}$ and is denoted by $\vec{D}*_M$.

The displacement vector concept is objective if and only if $\vec{D}_M = \vec{D}*_M$ in $E_3$. This equality can be obtained by using formulas (16)-(20).

In the description of $O$, the concept of the components of the deformation at the material point $P_{M_O}$ at the moment of time $M$ (i.e. at the moment $t_M = t$) are :

$$\gamma_{jk} = \frac{1}{2} \cdot ( \frac{\partial U_j}{\partial X_k} + \frac{\partial U_k}{\partial X_j} + \sum_{i=1}^{3} \frac{\partial U_i}{\partial X_k} \cdot \frac{\partial U_i}{\partial X_j} ) \qquad (21)$$

Components $\gamma_{jk}$, obtained using observer $O$ description, are related to components $\gamma*_{jk}$, obtained using observer $O*$ description, by relations

$$\gamma*_{jk} = \sum_{r=1}^{3} \sum_{q=1}^{3} \gamma_{rq} a_{jr} a_{kq} \qquad (22)$$

Relations (22) permit the definition of a concept (related to the deformation) independent on observer which is called the Cauchy deformation tensor of the body. That is the tensor which components in $R_O$ are given by (21).

### 3.3. Objectivity of the concept of the Cauchy small deformation tensor

In case of small displacements the deformation tensor components, in the description of $O$, can be approximated by :

$$\varepsilon_{jk} = \frac{1}{2} \cdot \left( \frac{\partial U_j}{\partial X_k} + \frac{\partial U_k}{\partial X_j} \right) \qquad (23)$$

and that in description of $O*$ by

$$\varepsilon*_{jk} = \frac{1}{2} \cdot \left( \frac{\partial U*_j}{\partial X*_k} + \frac{\partial U*_k}{\partial X*_j} \right) \qquad (24)$$

The components $\varepsilon_{jk}$ (obtained using observer $O$ description) are related to components $\varepsilon^*{}_{jk}$ (obtained using observer $O^*$ description) by relations

$$\varepsilon^*{}_{jk} = \sum_{r=1}^{3}\sum_{q=1}^{3}\varepsilon_{rq}a_{jr}a_{kq} \qquad (25)$$

Relations (25) permit the definition of the concept (independent on observer) called the Cauchy small deformation tensor of the body which components in $R_O$ are given by (23).

### 3.4. Objectivity of the concept of Cauchy stress tensor

In the description of observer $O$ the surface forces acting per unit area, in a material body $B$ at a moment of time $M$, is a vector valued function, called the stress vector, and is of the form [3]:

$$\vec{T} = \vec{T}(t_M, X_1, X_2, X_3; \vec{n}) \qquad (26)$$

Here: $\vec{T}$ represents the force at the moment of time $M$ (i.e. $t_M$) acting on the unit surface which passes through the material point of coordinate $(X_1, X_2, X_3)$ and which unit normal in that point is $\vec{n}$. The dependence of the stress vector $\vec{T}(t_M, X_1, X_2, X_3; \vec{n})$ on $\vec{n}$ according to [3] is given by:

$$\vec{T}(t_M, X_1, X_2, X_3; \vec{n}) = \sum_{i=1}^{3}\left(\sum_{j=1}^{3}\sigma_{ij}(t_M, X_1, X_2, X_3)\cdot n_j\right)\vec{e}_i \qquad (27)$$

where the coefficients $\sigma_{ij}(t_M, X_1, X_2, X_3)$ $i, j = 1,2,3$ are the coefficients of the Cauchy stress vector and $n_j$ are the components of the unit normal $\vec{n}$ with respect to $R_O$.

In the description of observer $O^*$ the surface forces acting per unit area, in the same material body $B$, at a moment of time $M$, is the stress vector:

$$\vec{T}^* = \vec{T}^*(t_M{}^*, X^*{}_1, X^*{}_2, X^*{}_3; \vec{n}) \qquad (28)$$

The dependence of $\vec{T}^*(t_M{}^*, X^*{}_1, X^*{}_2, X^*{}_3; \vec{n})$ on $\vec{n}$ is given by:

$$\vec{T}^*(t_M{}^*, X^*{}_1, X^*{}_2, X^*{}_3; \vec{n}) = \sum_{i=1}^{3}\left(\sum_{j=1}^{3}\sigma^*{}_{ij}(t_M{}^*, X^*{}_1, X^*{}_2, X^*{}_3)\cdot n^*{}_j\right)\vec{e}^*{}_i \qquad (29)$$

where the coefficients $\sigma^*{}_{ij}(t^*, X^*{}_1, X^*{}_2, X^*{}_3)$ $i, j = 1,2,3$ are the coefficients of the Cauchy stress vector and $n^*{}_j$ are the components of the unit normal $\vec{n}$ to the surface in the point of coordinate $(X^*{}_1, X^*{}_2, X^*{}_3)$ with respect to $R_{O^*}$.

The coefficients $\sigma_{jk}$ of the Cauchy stress vector, obtained using the observer $O$ description, are related to the coefficients $\sigma^*_{jk}$ of Cauchy stress vector, obtained using the observer $O^*$ description, by relations

$$\sigma^*_{jk} = \sum_{r=1}^{3}\sum_{q=1}^{3} \sigma_{rq} a_{jr} a_{kq} \qquad (30)$$

Relations (30) reconcile the meaning of concepts of Cauchy stress vector components in the description made by the two observers and permit the definition of a concept independent on observer which is called the Cauchy stress tensor of the body. The components in $R_0$ of this tensor are defined by (27)

### 3.5. Objectivity of the Hooke constitutive law

In case of a homogeneous and isotropic material body, the law oh Hooke concerning small deformations , in terms of the observer $O$, according to [3], is:

$$\sigma_{ij}(t, X_1, X_2, X_3) = \lambda \cdot \theta \cdot \delta_{ij} + 2\mu \cdot \varepsilon_{ij} \qquad (31)$$

where $\lambda$ and $\mu$ are the Lame constants $\delta_{ij}$ are the Kronecker coefficients and $\theta = \sum_{i=1}^{3} \varepsilon_{ii}$.

It turns that equation (31) is independent on observer (it is objective) and is called the constitutive law of Hooke in case of small deformations.

### 4. Caputo and Riemann-Liouville fractional order derivatives

The Caputo fractional order derivative was introduced by M. Caputo in 1967 [4]. According to [5] for a continuously differentiable function $f:[0,\infty) \to R$ the Caputo fractional derivative of order $\alpha$, $(0 < \alpha < 1)$ is defined by :

$$D_C^\alpha f(t) = \frac{1}{\Gamma(1-\alpha)} \cdot \int_0^t \frac{f'(\tau)}{(t-\tau)^\alpha} d\tau \qquad (32).$$

where $\Gamma$ is the Euler gamma function.

For a continuous function $f:[0,\infty) \to R$ the Riemann-Liouville fractional derivative of order $\alpha$, $(0 < \alpha < 1)$, according to [5], is defined by :

$$D_{R-L}^\alpha f(t) = \frac{1}{\Gamma(1-\alpha)} \cdot \frac{d}{dt} \int_0^t \frac{f(\tau)}{(t-\tau)^\alpha} d\tau \qquad (33)$$

Formulas (32) and (33) extends the first order derivative to fractional order derivatives.

## 5. Objectivity lost when Riemann-Liouville or Caputo fractional order derivatives are used.

In the context of the objectivity of the description of a mechanical event discussed in the present paper, in this section we show that objectivity is lost when in the description Riemann-Liouwille or Caputo fractional derivatives are used instead of the first and second order derivatives.

### 5.1. Objectivity is lost when velocity is defined using Riemann-Liouville fractional order derivatives

For that we consider a material point $P$ which moves on a line $L$ in $E_3$ and the movement is observed by two observers $O$ and $O^*$ placed in two different geometrical points $O$ and $O^*$ on the line. The reference frame of $O$ is $R_O = (O; \vec{e}_1, \vec{e}_2, \vec{e}_3)$, and that of $O^*$ is $R_{O^*} = (O^*; \vec{e}^*_1, \vec{e}^*_2, \vec{e}^*_3)$. Assume that the support in $E_3$ of the vectors $\vec{e}_1$ and $\vec{e}^*_1$ is the line L and $a_{11} = <\vec{e}^*_1, \vec{e}_1> = 1$. Each observer has his own chronometer which is started at the moment of time $M_O$ and $M_{O^*}$, when the material point $P$ passes for the first time in front of the observer $O$ and $O^*$ respectively. In case of the observer $O$ a moment of time $M$ is described by the real number $t_M$ and in case of the observer $O^*$ by the real number $t^*_M$. Remember that for the numbers $t_M$ and $t^*_M$ the following relations hold:

$$t_M = t^*_M + t_{M_{O^*}} \qquad\qquad t^*_M = t_M + t^*_{M_O}.$$

In the above relations $t_{M_{O^*}}$ is the real number which represent the moment $M_{O^*}$ in the system of time measuring of the observer $O$, and $t^*_{M_O}$ is the real number which represent the moment $M_O$ in the system of time measuring of the observer $O^*$. For precise the situation, assume that $0 = t_{M_O} < t_{M_{O^*}} < t_M$. For the coordinates of the material point $P$ in the representation of the observers $O$ and $O^*$ the following equalities hold:

$X_1(t_M) = X_{1O^*} + X^*_1(t^*_M), X_2(t_M) = 0, X_3(t_M) = 0$; or $X^*_1(t^*_M) = X^*_{1O} + X_1(t_M)$,
$X^*_2(t^*_M) = 0, X^*_3(t^*_M) = 0$.

Using the observer $O$ representation, the first component of the fractional order velocity, according to the Riemann-Liouville fractional order derivative of order $\alpha$, $(0 < \alpha < 1)$ is

$$D_{R-L}^\alpha X_1(t_M) = \frac{1}{\Gamma(1-\alpha)} \cdot \frac{d}{dt_M} \int_0^{t_M} \frac{X_1(\tau)}{(t_M - \tau)^\alpha} d\tau$$ while the second and the third components are equal to zero.

Using the observer $O^*$ representation, the first component of the fractional order velocity, according to the Riemann-Liouville fractional order derivative of order $\alpha$, $(0 < \alpha < 1)$ is

$$D_{R-L}^\alpha X^*_1(t^*_M) = \frac{1}{\Gamma(1-\alpha)} \cdot \frac{d}{dt^*_M} \int_0^{t^*_M} \frac{X^*_1(\xi)}{(t^*_M - \xi)^\alpha} d\xi$$ while the second and the third components

are equal to zero.

The fractional order velocity defined with Riemann-Liouville fractional order derivative $\alpha$, $(0 < \alpha < 1)$ is objective if and only if the following equality holds:

$$D_{R-L}{}^\alpha X_1(t_M) = D_{R-L}{}^\alpha X^*{}_1(t^*{}_M) \ .$$

Or using equalities $t_M = t^*{}_M + t_{M_{O^*}}$ and $X_1(t_M) = X_{1O^*} + X^*{}_1(t^*{}_M)$ it is easy to see that we have

$$D_{R-L}{}^\alpha X_1(t_M) = \frac{1}{\Gamma(1-\alpha)} \cdot \frac{d}{dt_M} \int_0^{t_{M_{O^*}}} \frac{X_1(\tau)}{(t_M - \tau)^\alpha} d\tau + D_{R-L}{}^\alpha X^*{}_1(t^*{}_M).$$

Hence, the fractional order velocity, defined with Riemann-Liouville fractional order derivative $\alpha$, $(0 < \alpha < 1)$ is objective, if and only if the following equality hold:

$$\frac{1}{\Gamma(1-\alpha)} \cdot \frac{d}{dt_M} \int_0^{t_{M_{O^*}}} \frac{X_1(\tau)}{(t_M - \tau)^\alpha} d\tau = 0$$

Because this equality in general is not fulfilled the velocity defined by using Riemann-Liouville fractional order derivative $\alpha$, $(0 < \alpha < 1)$ is not objective.

### 5.2. Objectivity is lost when fractional order velocity is defined by using Caputo fractional order derivatives.

Using the observer $O$ representation, the first component of the fractional order velocity, according to the Caputo fractional order derivative $\alpha$, $(0 < \alpha < 1)$ is

$$D_C{}^\alpha X_1(t_M) = \frac{1}{\Gamma(1-\alpha)} \cdot \int_0^{t_M} \frac{X'{}_1(\tau)}{(t_M - \tau)^\alpha} d\tau \quad \text{while the second and the third components are equal to zero.}$$

Using the observer $O^*$ representation the first component of the fractional order velocity according to the Caputo fractional order derivative of order $\alpha$, $(0 < \alpha < 1)$ is

$$D_C{}^\alpha X^*{}_1(t^*{}_M) = \frac{1}{\Gamma(1-\alpha)} \cdot \int_0^{t^*{}_M} \frac{X_1^{*'}(\xi)}{(t^*{}_M - \xi)^\alpha} d\xi \quad \text{while the second and the third components are equal}$$

to zero. The fractional order velocity defined with Caputo fractional order derivative of order $\alpha$, $(0 < \alpha < 1)$ is objective if and only if the following equality holds:

$$D_C{}^\alpha X_1(t_M) = D_C{}^\alpha X^*{}_1(t^*{}_M) \ .$$

Or using equalities $t_M = t^*{}_M + t_{M_{O^*}}$ and $X_1(t_M) = X_{1O^*} + X^*{}_1(t^*{}_M)$ it is easy to see that we have:

$$D_C{}^\alpha X_1(t_M) = \frac{1}{\Gamma(1-\alpha)} \cdot \int_0^{t_{M_{O^*}}} \frac{X'{}_1(\tau)}{(t_M - \tau)^\alpha} d\tau + D_C{}^\alpha X^*{}_1(t^*{}_M)$$

Hence, the fractional order velocity defined with Caputo fractional order derivative of order $\alpha$, $(0 < \alpha < 1)$ is objective if and only if the following equality holds:

$$\frac{1}{\Gamma(1-\alpha)} \cdot \int_0^{t_{M_{o^*}}} \frac{X'_1(\tau)}{(t_M - \tau)^\alpha} d\tau = 0$$

Because this equality in general is not fulfilled the velocity defined by using Caputo fractional order derivative $\alpha$, $(0 < \alpha < 1)$ is not objective.

*5.3. Objectivity is lost when fractional order acceleration is defined by using Riemann-Liouville fractional order derivatives $\alpha$ $(1 < \alpha < 2)$*

In the case of the considered example the fractional order acceleration defined with Riemann-Liouville fractional order derivative $\alpha$, $(1 < \alpha < 2)$ is objective if and only if the following equality holds:

$$D_{R-L}^\alpha X_1(t_M) = D_{R-L}^\alpha X^*_1(t^*_M)$$

Or in the case of the considered example the following equality holds:

$$D_{R-L}^\alpha X_1(t_M) = \frac{1}{\Gamma(2-\alpha)} \cdot \frac{d^2}{dt^2_M} \int_0^{t_{M_{o^*}}} \frac{X_1(\tau)}{(t_M - \tau)^{\alpha-1}} d\tau + D_{R-L}^\alpha X^*_1(t^*_M) \qquad (34)$$

Hence, the fractional order acceleration, defined with Riemann-Liouville fractional order derivative $\alpha$, $(1 < \alpha < 2)$ is objective, if and only if the following equality hold:

$$\frac{1}{\Gamma(2-\alpha)} \cdot \frac{d^2}{dt^2_M} \int_0^{t_{M_{o^*}}} \frac{X_1(\tau)}{(t_M - \tau)^{\alpha-1}} d\tau = 0$$

Because this equality in general is not fulfilled the acceleration defined by using Riemann-Liouville fractional order derivative $\alpha$, $(1 < \alpha < 2)$ is not objective.

*5.4. Objectivity is lost when fractional order acceleration is defined by using Caputo fractional order derivatives $\alpha$ $(1 < \alpha < 2)$.*

In the case of the considered example the fractional order acceleration defined with Caputo fractional order derivative $\alpha$, $(1 < \alpha < 2)$ is objective if and only if the following equality holds:

$$D_C^\alpha X_1(t_M) = D_C^\alpha X^*_1(t^*_M)$$

Or in the case of the considered example the following equality holds:

$$D_C^\alpha X_1(t_M) = \frac{1}{\Gamma(2-\alpha)} \cdot \int_0^{t_{M_{o^*}}} \frac{X''_1(\tau)}{(t_M - \tau)^{\alpha-1}} d\tau + D_C^\alpha X^*_1(t^*_M) \qquad (35)$$

Hence, the fractional order acceleration, defined with Caputo fractional order derivative $\alpha$, $(1 < \alpha < 2)$ is objective, if and only if the following equality hold:

$$\frac{1}{\Gamma(2-\alpha)} \cdot \int_0^{t_{M_o*}} \frac{X''_1(\tau)}{(t_M - \tau)^{\alpha-1}} d\tau = 0$$

Because this equality in general is not fulfilled the acceleration defined by using Caputo fractional order derivative $\alpha$, $(1 < \alpha < 2)$ is not objective.

*5.5. Objectivity is lost when in the second law of Newton the acceleration is defined by using Riemann-Liouville fractional order derivatives $\alpha$, $(1 < \alpha < 2)$.*

In terms of the description of observer $O$, the second law of Newton, leads to the conclusion that the functions $X_1(t_M)$, $X_2(t_M)$, $X_3(t_M)$, describing the motion under the action of a force field which acts along the line $L$ and depends only on the position, satisfy the following system of differential equations:

$$mD_{R-L}^{\alpha} X_1(t_M) = F_{1O}(X_1(t_M)) \ , \ mD_{R-L}^{\alpha} X_2(t_M) = 0 \ , \ mD_{R-L}^{\alpha} X_3(t_M) = 0 \qquad (36)$$

In terms of the description of observer $O^*$, the second law of Newton leads to the conclusion that the functions $X^*_1(t^*_M)$, $X^*_2(t^*_M)$, $X^*_3(t^*_M)$, describing the motion under the action of the same force field, satisfy the following system of differential equations:

$$mD_{R-L}^{\alpha} X^*_1(t^*_M) = F^*_{1O}(X^*_1(t^*_M)), \ mD_{R-L}^{\alpha} X^*_2(t^*_M) = 0, \ mD_{R-L}^{\alpha} X^*_3(t^*_M) = 0 \quad (37)$$

Equalities $mD_{R-L}^{\alpha} X_1(t_M) = F_{1O}(X_1(t_M))$ , $mD_{R-L}^{\alpha} X^*_1(t^*_M) = F^*_{1O}(X^*_1(t^*_M))$ and (34) implies:

$$F_{1O}(X_1(t_M)) = F^*_{1O}(X^*_1(t^*_M)) + \frac{m}{\Gamma(2-\alpha)} \cdot \frac{d^2}{dt^2_M} \int_0^{t_{M_o*}} \frac{X_1(\tau)}{(t_M - \tau)^{\alpha-1}} d\tau \qquad (38)$$

from where it follows that the following equality holds:

$$\frac{m}{\Gamma(2-\alpha)} \cdot \frac{d^2}{dt^2_M} \int_0^{t_{M_o*}} \frac{X_1(\tau)}{(t_M - \tau)^{\alpha-1}} d\tau = 0 \qquad (39)$$

But in general equality (39) is not valid. For this reason the description of dynamic with the system of differential equations (36) is not objective.

*5.6. Objectivity is lost when in the second law of Newton Caputo fractional order derivatives $\alpha$,*

*$(1 < \alpha < 2)$ are used*

In terms of the description of observer $O$, the second law of Newton, leads to the conclusion that the functions $X_1(t_M)$, $X_2(t_M)$, $X_3(t_M)$, describing the motion under the action of a force field which acts

along the line $L$ and depends only on the position, satisfy the following system of differential equations:

$$mD_C^\alpha X_1(t_M) = F_{1O}(X_1(t_M)) \, , \; mD_C^\alpha X_2(t_M) = 0 \, , \; mD_C^\alpha X_3(t_M) = 0 \tag{40}$$

In terms of the description of observer $O^*$, the second law of Newton, leads to the conclusion that the functions $X^*_1(t^*_M), X^*_2(t^*_M), X^*_3(t^*_M)$, describing the motion under the action of the same force field, satisfy the following system of differential equations:

$$mD_C^\alpha X^*_1(t^*_M) = F^*_{1O}(X^*_1(t^*_M)), \; mD_C^\alpha X^*_2(t^*_M) = 0 \, , \; mD_C^\alpha X^*_3(t^*_M) = 0 \tag{41}$$

Equalities $mD_C^\alpha X_1(t_M) = F_{1O}(X_1(t_M))$ , $mD_C^\alpha X^*_1(t^*_M) = F^*_{1O}(X^*_1(t^*_M))$ and (35) imply

$$F_{1O}(X_1(t_M)) = F^*_{1O}(X^*_1(t^*_M)) + \frac{m}{\Gamma(2-\alpha)} \cdot \int_0^{t_{M_{o^*}}} \frac{X''_1(\tau)}{(t_M - \tau)^{\alpha-1}} d\tau \tag{42}$$

from where:

$$\frac{m}{\Gamma(2-\alpha)} \cdot \int_0^{t_{M_{o^*}}} \frac{X''_1(\tau)}{(t_M - \tau)^{\alpha-1}} d\tau = 0 \tag{43}$$

This last equality in general is not valid. For this reason the description of dynamic with the system of differential equations (40) is not objective.

### 5.7. Objectivity is lost when in the constitutive law of Hooke Riemann-Liouville fractional order derivatives $\alpha, (0 < \alpha < 1)$ are used

In [9] Bagley and Torvik instead of the constitutive law of Hooke (31) consider the constitutive law:

$$\sigma_{ij}(t, X_1, X_2, X_3) = \lambda \cdot \theta(t) \cdot \delta_{ij} + 2\mu \cdot D_{R-L}^\alpha \varepsilon_{ij}(t) \tag{44}$$

If (44) is the Hooke constitutive law in the observer $O$ description and the description is objective, then the constitutive law in the observer $O^*$ description is

$$\sigma^*_{ij}(t^*, X^*_1, X^*_2, X^*_3) = \lambda \cdot \theta^*(t) \cdot \delta_{ij} + 2\mu \cdot D_{R-L}^\alpha \varepsilon^*_{ij}(t^*) \tag{45}$$

In order to see that (44) is not objective start with (44) consider $i$ different from $j$, use the equality

$$D_{R-L}^\alpha \varepsilon_{ij}(t_M) = \frac{1}{\Gamma(1-\alpha)} \cdot \frac{d}{dt_M} \int_0^{t_{M_{o^*}}} \frac{\varepsilon_{ij}(\tau)}{(t_M - \tau)^\alpha} d\tau + D_{R-L}^\alpha \varepsilon^*_{ij}(t^*_M) \tag{46}$$

and obtain the following equality:

$$\sigma^*_{ij}(t^*, X^*_1, X^*_2, X^*_3) = 2\mu \cdot D_{R-L}^\alpha \varepsilon^*_{ij}(t^*) + \frac{2 \cdot \mu}{\Gamma(1-\alpha)} \sum_{k=1}^{3}\sum_{l=1}^{3} a_{ki}a_{lj} \cdot \frac{d}{dt_M} \int_0^{t_{M_{o^*}}} \frac{\varepsilon_{ij}(\tau)}{(t_M - \tau)^\alpha} d\tau$$

Hence, by using (45) it follows that if the description (44) is objective then the following equality holds:

$$\frac{2\cdot\mu}{\Gamma(1-\alpha)}\sum_{k=1}^{3}\sum_{l=1}^{3}a_{ki}a_{lj}\cdot\frac{d}{dt_M}\int_{0}^{t_{M_{o^*}}}\frac{\varepsilon_{ij}(\tau)}{(t_M-\tau)^\alpha}d\tau=0$$

Because this equality in general is not fulfilled the constitutive law described with (44) is not objective.

**6. Final remark**

In the scientific literature there exists a lot of papers using fractional order derivatives for describing real world phenomena. For example, in [10] the authors use a fractional conservation of mass equation to model fluid flow when the control volume is not large enough compared to the scale of heterogeneity and when the flux within the control volume is non-linear; in [11], [12] Atangana et al. described some groundwater flow problems using the concept of derivative with fractional order; the authors of [13]-[15] use fractional order differential equations for modeling contaminant flow in heterogeneous porous media; for describing anomalous diffusion processes in complex media fractional-order diffusion equation models are considered in [16] and [17]; in [6],[7],[9], [18] the authors uses fractional order derivatives in linear viscoelasticity; in [19]-[21] the authors use fractional order derivatives for acoustical wave propagation in complex media; in [22], [23] the fractional order Schrodinger equation and the variable –order fractional Schrodinger equation are considered. It would be interesting to know if these descriptions are objective or not.

**7. Conclusion**

Beside the conclusion which is the title of the paper the main conclusion of this study is in fact the following question: if a mathematical description of a real phenomenon is not objective then what is the interpretation of the reported results and how this results have to be used?